\documentstyle[sprocl]{article}

\def\be{\begin{equation}}
\def\ee{\end{equation}}
\def\bea{\begin{eqnarray}}
\def\eea{\end{eqnarray}}
\def\etal{{\it et al.}}
\def\lesssim{\mathrel{\hbox{\rlap{\hbox{\lower4pt\hbox{$\sim$}}}\hbox{$<$}}}}
\def\gtrsim{\mathrel{\hbox{\rlap{\hbox{\lower4pt\hbox{$\sim$}}}\hbox{$>$}}}}

\begin{document}

\title{RESULTS FROM THE HIGH-Z SUPERNOVA SEARCH TEAM}

\author{ALEXEI V. FILIPPENKO and ADAM G. RIESS}

\address{Department of Astronomy, University of California, \\
Berkeley, California 94720-3411, USA}

\author {(on behalf of the High-Z Supernova Search Team)}

\bigskip


\maketitle\abstracts{ABSTRACT: We review the use of Type Ia supernovae
for cosmological distance determinations.  Low-redshift SNe~Ia ($z
\lesssim 0.1$) demonstrate that (a) the Hubble expansion is linear,
(b) $H_0 = 65 \pm 2$ (statistical) km s$^{-1}$ Mpc$^{-1}$, (c) the
bulk motion of the Local Group is consistent with the COBE result, and
(d) the properties of dust in other galaxies are similar to those of
dust in the Milky Way.  We find that the light curves of high-redshift
SNe~Ia are stretched in a manner consistent with the expansion of
space; similarly, their spectra exhibit slower temporal evolution (by
a factor of $1 + z$) than those of nearby SNe~Ia.  The luminosity
distances of our 16 high-redshift SNe~Ia are, on average, 10--15\%
farther than expected in a low mass-density ($\Omega_M=0.2$) universe
without a cosmological constant.  Our analysis strongly supports
eternally expanding models with positive cosmological constant and a
current acceleration of the expansion.  We address many potential
sources of systematic error; at present, none of them reconciles the
data with $\Omega_\Lambda=0$ and $q_0 \geq 0$.  The dynamical age of
the Universe is estimated to be $14.2 \pm 1.7$ Gyr, consistent with
the ages of globular star clusters.}

\section{Introduction}

    Supernovae (SNe) come in two main varieties (see Filippenko 1997b
for a review). Those whose optical spectra exhibit hydrogen are
classified as Type II, while hydrogen-deficient SNe are designated
Type I. SNe~I are further subdivided according to the appearance of
the early-time spectrum: SNe~Ia are characterized by strong absorption
near 6150~\AA\ (now attributed to Si~II), SNe~Ib lack this feature but
instead show prominent He~I lines, and SNe~Ic have neither the Si~II
nor the He~I lines. SNe~Ia are believed to result from the
thermonuclear disruption of carbon-oxygen white dwarfs, while SNe~II
come from core collapse in massive supergiant stars. The latter
mechanism probably produces most SNe~Ib/Ic as well, but the progenitor
stars previously lost their outer layers of hydrogen or even helium.

   It has long been recognized that SNe~Ia may be very useful distance
indicators for a number of reasons (Branch \& Tammann 1992; Branch
1998, and references therein). (1) They are exceedingly luminous, with
peak absolute blue magnitudes averaging $-19.2$ if the Hubble
constant, $H_0$, is 65 km s$^{-1}$ Mpc$^{-1}$. (2) ``Normal" SNe~Ia
have small dispersion among their peak absolute magnitudes ($\sigma
\lesssim 0.3$ mag). (3) Our understanding of the progenitors and
explosion mechanism of SNe~Ia is on a reasonably firm physical basis.
(4) Little cosmic evolution is expected in the peak luminosities of
SNe~Ia, and it can be modeled. This makes SNe~Ia superior to
galaxies as distance indicators. (5) One can perform {\it local} tests
of various possible complications and evolutionary effects by
comparing nearby SNe~Ia in different environments.

   Research on SNe~Ia in the 1990s has demonstrated their enormous
potential as cosmological distance indicators. Although there are
subtle effects that must indeed be taken into account, it appears that
SNe~Ia provide among the most accurate values of $H_0$, $q_0$ (the
deceleration parameter), $\Omega_M$ (the matter density), and
$\Omega_\Lambda$ (the cosmological constant, $c^2\Lambda/3H_0^2$).

   There are now two major teams involved in the systematic
investigation of high-redshift SNe~Ia for cosmological purposes. The
``Supernova Cosmology Project" (SCP) is led by Saul Perlmutter of the
Lawrence Berkeley Laboratory, while the ``High-Z Supernova Search
Team" (HZT) is led by Brian Schmidt of the Mt. Stromlo and Siding
Springs Observatories. One of us (A.V.F.) has worked with both teams,
although his primary allegiance is now with the HZT. In this lecture
we present results from the HZT.

\section{Homogeneity and Heterogeneity}

   The traditional way in which SNe~Ia have been used for cosmological
distance determinations has been to assume that they are perfect
``standard candles" and to compare their observed peak brightness with
those of SNe~Ia in galaxies whose distances have been independently
determined (e.g., Cepheids). The rationale is that SNe~Ia exhibit
relatively little scatter in their peak blue luminosity ($\sigma_B
\approx 0.4$--0.5 mag; Branch \& Miller 1993), and even less if
``peculiar" or highly reddened objects are eliminated from
consideration by using a color cut.  Moreover, the optical spectra of
SNe~Ia are usually quite homogeneous, if care is taken to compare
objects at similar times relative to maximum brightness (Riess \etal
~1997, and references therein). Branch \etal ~(1993) estimate that over
80\% of all SNe~Ia discovered thus far are ``normal."

   From a Hubble diagram constructed with unreddened, moderately
distant SNe~Ia ($z \lesssim 0.1$) for which peculiar motions should be
small and relative distances (as given by ratios of redshifts) are
accurate, Vaughan \etal ~(1995) find that

\[ <M_B({\rm max})> = (-19.74 \pm 0.06) + 5\, {\rm
log}\, (H_0/50)~{\rm mag}. \]

\noindent In a series of papers, Sandage \etal ~(1996) and Saha \etal
~(1996) combine similar relations with {\it Hubble Space Telescope
(HST)} Cepheid distances to the host galaxies of six SNe~Ia to derive
$H_0 = 57 \pm 4$ km s$^{-1}$ Mpc$^{-1}$.

   Over the past decade it has become clear, however, that SNe~Ia do
{\it not} constitute a perfectly homogeneous subclass (e.g.,
Filippenko 1997a). In retrospect this should have been obvious: the
Hubble diagram for SNe~Ia exhibits scatter larger than the photometric
errors, the dispersion actually {\it rises} when reddening corrections
are applied (under the assumption that all SNe~Ia have uniform, very
blue intrinsic colors at maximum; van den Bergh \& Pazder
1992; Sandage \& Tammann 1993), and there are some significant
outliers whose anomalous magnitudes cannot possibly be explained by
extinction alone.

    Spectroscopic and photometric peculiarities have been noted with
increasing frequency in well-observed SNe~Ia. A striking case is SN
1991T; its pre-maximum spectrum did not exhibit Si~II or Ca~II
absorption lines, yet two months past maximum the spectrum was nearly
indistinguishable from that of a classical SN~Ia (Filippenko \etal
~1992b; Phillips \etal ~1992).  The light curves of SN 1991T were
slightly broader than the SN~Ia template curves, and the object was
probably somewhat more luminous than average at maximum. The reigning
champion of well observed, peculiar SNe~Ia is SN 1991bg (Filippenko
\etal ~1992a; Leibundgut \etal ~1993; Turatto \etal ~1996). At maximum
brightness it was subluminous by 1.6 mag in $V$ and 2.5 mag in $B$,
its colors were intrinsically red, and its spectrum was peculiar (with
a deep absorption trough due to Ti~II).  Moreover, the decline from
maximum brightness was very steep, the $I$-band light curve did not
exhibit a secondary maximum like normal SNe~Ia, and the velocity of
the ejecta was unusually low. The photometric heterogeneity among
SNe~Ia is well demonstrated by Suntzeff (1996) with five objects
having excellent $BVRI$ light curves.

\section {Cosmological Uses}

\subsection {Luminosity Corrections and Nearby Supernovae}

   Although SNe~Ia can no longer be considered perfect ``standard
candles," they are still exceptionally useful for cosmological
distance determinations. Excluding those of low luminosity (which are
hard to find, especially at large distances), most SNe~Ia are {\it
nearly} standard (Branch \etal ~1993). Also, after many tenuous
suggestions, convincing evidence has finally been found for a {\it
correlation} between light-curve shape and luminosity. Phillips (1993)
achieved this by quantifying the photometric differences among a set
of nine well-observed SNe~Ia using a parameter, $\Delta m_{15}(B)$,
which measures the total drop (in $B$ magnitudes) from maximum to $t =
15$ days after $B$ maximum. In all cases the host galaxies of his
SNe~Ia have accurate relative distances from surface brightness
fluctuations or from the Tully-Fisher relation. In $B$, the SNe~Ia
exhibit a total spread of $\sim 2$ mag in maximum luminosity, and the
intrinsically bright SNe~Ia clearly decline more slowly than dim
ones. The range in absolute magnitude is smaller in $V$ and $I$,
making the correlation with $\Delta m_{15}(B)$ less steep than in $B$,
but it is present nonetheless.

   Using SNe~Ia discovered during the Cal\'an/Tololo survey ($z
\lesssim 0.1$), Hamuy \etal ~(1995, 1996b) confirm and refine the
Phillips (1993) correlation between $\Delta m_{15}(B)$ and
$M_{max}(B,V)$: it is not as steep as had been claimed. Apparently the
slope is steep only at low luminosities; thus, objects such as SN
1991bg skew the slope of the best-fitting single straight line. Hamuy
\etal ~reduce the scatter in the Hubble diagram of normal, unreddened
SNe~Ia to only 0.17 mag in $B$ and 0.14 mag in $V$; see also Tripp
(1997).

   In a similar effort, Riess \etal ~(1995a) show that the luminosity
of SNe~Ia correlates with the detailed shape of the light curve, not
just with its initial decline. They form a ``training set" of
light-curve shapes from 9 well-observed SNe~Ia having known relative
distances, including very peculiar objects (e.g., SN 1991bg). When the
light curves of an independent sample of 13 SNe~Ia (the Cal\'an/Tololo
survey) are analyzed with this set of basis vectors, the dispersion in
the $V$-band Hubble diagram drops from 0.50 to 0.21 mag, and the
Hubble constant rises from $53 \pm 11$ to $67 \pm 7$ km s$^{-1}$
Mpc$^{-1}$, comparable to the conclusions of Hamuy \etal ~(1995,
1996b). About half of the rise in $H_0$ results from a change in the
position of the ``ridge line" defining the linear Hubble relation, and
half is from a correction to the luminosity of some of the local
calibrators which appear to be unusually luminous (e.g., SN 1972E).

   By using light-curve shapes measured through several different
filters, Riess \etal ~(1996) extend their analysis and objectively
eliminate the effects of interstellar extinction: a SN~Ia that has an
unusually red $B-V$ color at maximum brightness is assumed to be {\it
intrinsically} subluminous if its light curves rise and decline
quickly, or of normal luminosity but significantly {\it reddened} if
its light curves rise and decline slowly. With a set of 20 SNe~Ia
consisting of the Cal\'an/Tololo sample and their own objects, Riess
\etal ~(1996) show that the dispersion decreases from 0.52 mag to 0.12
mag after application of this ``multi-color light curve shape" (MLCS)
method. Preliminary results with a very recent, expanded set of nearly
50 SNe~Ia indicate that the dispersion decreases from 0.44 mag to 0.15
mag (Riess \etal ~1998c). The resulting Hubble constant is $65 \pm 2$
(statistical) km s$^{-1}$ Mpc$^{-1}$, with an additional systematic
and zero-point uncertainty of $\pm 5$ km s$^{-1}$ Mpc$^{-1}$. Riess
\etal ~(1996) also show that the Hubble flow is remarkably linear;
indeed, SNe~Ia now constitute the best evidence for
linearity. Finally, they argue that the dust affecting SNe~Ia is {\it
not} of circumstellar origin, and show quantitatively that the
extinction curve in external galaxies typically does not differ from
that in the Milky Way (cf. Branch \& Tammann 1992; but see Tripp 
1998).

   Riess \etal ~(1995b) capitalize on another use of SNe~Ia:
determination of the Milky Way Galaxy's peculiar motion relative to
the Hubble flow. They select galaxies whose distances were accurately
determined from SNe~Ia, and compare their observed recession
velocities with those expected from the Hubble law alone. The speed
and direction of the Galaxy's motion are consistent with what is found
from COBE (Cosmic Background Explorer) studies of the microwave
background, but not with the results of Lauer \& Postman (1994).

   The advantage of systematically correcting the luminosities of
SNe~Ia at high redshifts rather than trying to isolate ``normal"
ones seems clear in view of recent evidence that the luminosity of
SNe~Ia may be a function of stellar population. If the most luminous
SNe~Ia occur in young stellar populations (Hamuy \etal ~1995,
1996a; Branch \etal ~1996a), then we might expect the mean peak
luminosity of high-redshift SNe~Ia to differ from that of a local
sample. Alternatively, the use of Cepheids (Population I objects) to
calibrate local SNe~Ia can lead to a zero point that is too luminous.
On the other hand, as long as the physics of SNe~Ia is essentially the
same in young stellar populations locally and at high redshift, we
should be able to adopt the luminosity correction methods (photometric 
and spectroscopic) found from detailed studies of nearby samples of
SNe~Ia.

\section{High-Redshift Supernovae}

\subsection{The Search}

   These same techniques can be applied to construct a Hubble diagram
with high-redshift SNe, from which the value of $q_0$ can be
determined. With enough objects spanning a range of redshifts, we can
determine $\Omega_M$ and $\Omega_\Lambda$ independently (e.g., Goobar
\& Perlmutter 1995). Contours of peak apparent $R$-band magnitude 
for SNe~Ia at two redshifts have different slopes in the
$\Omega_M$--$\Omega_\Lambda$ plane, and the regions of intersection
provide the answers we seek.

   Based on the pioneering work of Norgaard-Nielsen \etal ~(1989),
whose goal was to find SNe in moderate-redshift clusters of galaxies,
Perlmutter \etal ~(1997) and our team (Schmidt \etal ~1998) devised a
strategy that almost guarantees the discovery of many faint, distant
SNe~Ia on demand, during a predetermined set of nights.  This ``batch"
approach to studying distant SNe allows follow-up spectroscopy and
photometry to be {\it scheduled} in advance, resulting in a systematic
study not possible with random discoveries.  Most of the searched
fields are equatorial, permitting follow-up from both hemispheres.

    Our approach is simple in principle; see Schmidt \etal ~(1998) for
details, and for a description of our first high-redshift SN~Ia (SN
1995K). Pairs of first-epoch images are obtained with the CTIO or CFHT
4-m telescopes and wide-angle imaging cameras during the nights just
after new moon, followed by second-epoch images 3--4 weeks later.
(Pairs of images permit removal of cosmic rays, asteroids, and distant
Kuiper-belt objects.) These are compared immediately using well-tested
software, and new SN candidates are identified in the second-epoch
images. Spectra are obtained as soon as possible after discovery to
verify that the objects are SNe~Ia and determine their redshifts.
Each team has already found over 60 SNe in concentrated batches, as
reported in numerous IAU Circulars (e.g., Perlmutter \etal ~1995 ---
11 SNe with $0.16 \lesssim z \lesssim 0.65$; Suntzeff \etal ~1996 ---
17 SNe with $0.09 \lesssim z \lesssim 0.84$).

   Intensive photometry of the SNe~Ia commences within a few days
after procurement of the second-epoch images; it is continued
throughout the ensuing and subsequent dark runs. In a few cases {\it
HST} images are obtained. As expected, most of the discoveries are
{\it on the rise or near maximum brightness}.  When possible, the SNe
are observed in filters which closely match the redshifted $B$ and $V$
bands; this way, the $K$-corrections become only a second-order effect
(Kim \etal ~1996). Custom-designed filters for redshifts centered on
0.35 and 0.45 are used by our team (Schmidt \etal ~1998), when
appropriate.  We try to obtain excellent {\it multi-color} light
curves, so that reddening and luminosity corrections can be applied
(Riess \etal ~1996; Hamuy \etal ~1996a,b).

  Although SNe in the magnitude range 22--22.5 can sometimes be
spectroscopically confirmed with 4-m class telescopes, the
signal-to-noise ratios are low, even after several hours of
integration. Certainly Keck is required for the fainter objects (mag
22.5--24.5). With Keck, not only can we rapidly confirm a large number
of candidate SNe, but we can search for peculiarities in the spectra
that might indicate evolution of SNe~Ia with redshift.
Moreover, high-quality spectra allow us to measure the age of
a supernova: we have developed a method for automatically comparing
the spectrum of a SN~Ia with a library of spectra corresponding to
many different epochs in the development of SNe~Ia (Riess \etal
~1997). Our technique also has great practical utility at the
telescope: we can determine the age of a SN ``on the fly,'' within
half an hour after obtaining its spectrum. This allows us to rapidly
decide which SNe are the best ones for subsequent photometric
follow-up, and we immediately alert our collaborators on other
telescopes.

\subsection {Results}

   First, we note that the light curves of high-redshift SNe~Ia are
broader than those of nearby SNe~Ia; the initial indications of
Leibundgut \etal ~(1996) and Goldhaber \etal ~(1997) are amply
confirmed with our larger samples. Quantitatively, the amount by which
the light curves are ``stretched'' is consistent with a factor of $1 +
z$, as expected if redshifts are produced by the expansion of space
rather than by ``tired light.'' We were also able to demonstrate this
{\it spectroscopically} at the $2\sigma$ confidence level for a single
object: the spectrum of SN 1996bj ($z = 0.57$) evolved more slowly
than those of nearby SNe~Ia, by a factor consistent with $1 + z$
(Riess \etal ~1997). More recently, we have used observations of SN
1997ex ($z = 0.36$) at three epochs to conclusively verify the effects
of time dilation: temporal changes in the spectra are slower than
those of nearby SNe~Ia by roughly the expected factor of 1.36
(Filippenko \etal ~1998).

   Following our Spring 1997 campaign, in which we found a SN with $z
= 0.97$ (the highest known for a SN, but we do not have spectroscopic
confirmation that it is a SN~Ia), and for which we obtained {\it HST}
follow-up of three SNe, we published our first substantial results
concerning the density of the Universe (Garnavich \etal ~1998a):
$\Omega_M = 0.35 \pm 0.3$ under the {\it assumption} that $\Omega_{\rm
total} = 1$, or $\Omega_M = -0.1 \pm 0.5$ under the {\it assumption}
that $\Omega_\Lambda = 0$.  Our independent analysis of 10 SNe~Ia
using the ``snapshot'' distance method (with which conclusions are
drawn from sparsely observed SNe~Ia) gives quantitatively similar
conclusions (Riess \etal ~1998a).

   Our more recent results (Riess \etal ~1998b), obtained from a total
of 16 high-$z$ SNe~Ia, were announced for the first time at this
meeting. The Hubble diagram (from a refined version of the MLCS
method; Riess \etal ~1998b) for the 10 best-observed high-$z$ SNe~Ia is
given in Figure 1, while Figure 2 illustrates the derived confidence
contours in the $\Omega_M$--$\Omega_\Lambda$ plane. We confirm our
previous suggestion that $\Omega_M$ is low. Even more exciting,
however, is our conclusion that $\Omega_\Lambda$ is {\it nonzero} at
the 3$\sigma$ statistical confidence level. With the MLCS method
applied to the full set of 16 SNe~Ia, our formal results are $\Omega_M
= 0.24 \pm 0.10$ if $\Omega_{\rm total} = 1$, or $\Omega_M = -0.35 \pm
0.18$ (unphysical) if $\Omega_\Lambda = 0$. If we demand that
$\Omega_M = 0.2$, then the best value for $\Omega_\Lambda$ is $0.66
\pm 0.21$. [Our data do not provide assumption-free, independent
constraints on $\Omega_M$ and $\Omega_\Lambda$ to high precision
without ancillary assumptions or inclusion of a supernova (SN 1997ck)
whose classification and extinction are uncertain.]  These conclusions
do not change significantly if only the 10 best-observed SNe~Ia
(Fig. 1) are used.  The $\Delta m_{15}(B)$ method yields similar
results; if anything, the case for a positive cosmological constant
strengthens. (For brevity, in this paper we won't quote the $\Delta
m_{15}(B)$ numbers; see Riess \etal ~1998b.)  Note that the current
upper limit on $\Omega_\Lambda$ from studies of gravitational lenses
($\leq 0.66$ at 95\% confidence; Kochanek 1996) is not formally
inconsistent with our results.

   Though not drawn in Figure 2, the expected confidence contours from
measurements of the angular scale of the first Doppler peak of the
cosmic microwave background radiation (CMBR) are nearly perpendicular
to those provided by SNe~Ia (Zaldarriaga \etal ~1997); thus, the two
techniques provide complementary information.  The space-based CMBR
experiments in the next decade (e.g., MAP, Planck) will give very
narrow ellipses, but a stunning result is already provided by existing
measurements (Hancock \etal ~1998): analysis done by our team after
this meeting demonstrates that $\Omega_M + \Omega_\Lambda = 0.94 \pm
0.26$, when the SN and CMBR constraints are combined (Garnavich \etal
~1998b). The confidence contours are nearly circular, instead of highly
eccentric ellipses as in Figure 2.  We eagerly look forward to future
CMBR measurements of even greater precision.

   The dynamical age of the Universe can be calculated from the
cosmological parameters. In an empty Universe with no cosmological
constant, the dynamical age is simply the ``Hubble time" (i.e., the
inverse of the Hubble constant); there is no deceleration.  SNe~Ia
yield $H_0 = 65 \pm 2$ km s$^{-1}$ Mpc$^{-1}$ (statistical uncertainty
only), and a Hubble time of $15.1 \pm 0.5$ Gyr. For a more complex
cosmology, integrating the velocity of the expansion from the current
epoch ($z=0$) to the beginning ($z=\infty$) yields an expression for
the dynamical age. As shown in detail by Riess \etal ~(1998b), we
obtain a value of 14.2$^{+1.0}_{-0.8}$ Gyr using the likely range for
$(\Omega_M, \Omega_\Lambda)$ that we measure.  (The precision is so
high because our experiment is sensitive to the {\it difference}
between $\Omega_M$ and $\Omega_\Lambda$, and the dynamical age also
varies approximately this way.)  Including the {\it systematic}
uncertainty of the Cepheid distance scale, which may be up to 10\%, a
reasonable estimate of the dynamical age is $14.2 \pm 1.7$ Gyr.

  This result is consistent with ages determined from various other
techniques such as the cooling of white dwarfs (Galactic disk $> 9.5$
Gyr; Oswalt \etal ~1996), radioactive dating of stars via the thorium
and europium abundances ($15.2 \pm 3.7$ Gyr; Cowan \etal ~1997), and
studies of globular clusters (10--15 Gyr, depending on whether {\it
Hipparcos} parallaxes of Cepheids are adopted; Gratton \etal ~1997;
Chaboyer \etal ~1998).  Evidently, there is no longer a problem that
the age of the oldest stars is greater than the dynamical age of the
Universe.

\section{Discussion}

   {\it High-redshift SNe Ia are observed to be dimmer than expected
in an empty Universe (i.e., $\Omega_M=0$) with no cosmological
constant.}  A cosmological explanation for this observation is that a
positive vacuum energy density accelerates the expansion.  Mass
density in the Universe exacerbates this problem, requiring even more
vacuum energy.  For a Universe with $\Omega_M=0.2$, the average MLCS
distance moduli of the well-observed SNe are 0.25 mag larger (i.e.,
12.5\% greater distances) than the prediction from $\Omega_\Lambda=0$.
The average MLCS distance moduli are still 0.18 mag bigger than
required for a 68.3\% (1$\sigma$) consistency for a Universe with
$\Omega_M=0.2$ and without a cosmological constant.  The derived value
of $q_0$ is $-0.75 \pm 0.32$, implying that the expansion of the
Universe is accelerating.  It appears that the Universe will expand
eternally.

\subsection {Comparisons}

   The results given here are consistent with other reported
observations of high-redshift SNe~Ia from our HZT (Garnavich \etal
~1998a; Schmidt \etal ~1998), and the improved statistics of this larger
sample reveal the potential influence of a positive cosmological
constant.

     Our results are {\it inconsistent} at the $\sim 2\sigma$
confidence level with those of Perlmutter \etal ~(1997), who found
$\Omega_M=0.94 \pm 0.3 \, (\Omega_\Lambda=0.06)$ for a flat Universe
and $\Omega_M=0.88 \pm 0.64$ for $\Omega_\Lambda \equiv 0$.  They are
marginally consistent with those of Perlmutter \etal ~(1998) who, with
the addition of a single very high-redshift SN~Ia ($z=0.83$), found
$\Omega_M=0.6 \pm 0.2 \, (\Omega_\Lambda=0.4)$ for a flat Universe and
$\Omega_M=0.2 \pm 0.4$ for $\Omega_\Lambda \equiv 0$. They are
consistent with the results presented at this meeting by Perlmutter
and Goldhaber: they stated that analysis of 40 SNe~Ia yields
$\Omega_M=0.25 \pm 0.06 \, (\Omega_\Lambda=0.75)$ for a flat Universe
and $\Omega_M=-0.4 \pm 0.1$ for $\Omega_\Lambda \equiv 0$. It will
be of interest to learn the reason for the monotonic shift in
the results of Perlmutter's SCP.

  Although the HZT experiment reported here is very similar to that
performed by the SCP (Perlmutter \etal ~1997, 1998), there are some
differences worth noting. The HZT explicitly corrects for the effects
of extinction evidenced by reddening of the SN~Ia colors (Schmidt
\etal ~1998; Garnavich \etal ~1998a; Riess \etal ~1998b). Not correcting
for extinction in the nearby and distant samples could affect the
cosmological results in either direction since we do not know the sign
of the difference of the mean extinction. We also include $H_0$ 
as a free parameter in each of our fits to the other cosmological
parameters.  Treating the nearby sample in the same way as the 
distant sample is a crucial requirement of this work.  Our experience
observing the nearby sample aids our ability to accomplish this goal.

\subsection{Systematic Effects} 

   A very important point is that the dispersion in the peak
luminosities of SNe~Ia ($\sigma = 0.15$ mag) is low after application
of the MLCS method of Riess \etal ~(1996, 1998b). With 16 SNe~Ia, our
effective uncertainty is $0.15/4 \approx 0.04$ mag, less than the
expected difference of 0.25 mag between universes with $\Omega_\Lambda
= 0$ and 0.76 (and low $\Omega_M$); see Figure 1.  Systematic
uncertainties of even 0.05 mag (e.g., in the extinction) are
significant, and at 0.1 mag they dominate any decrease in statistical
uncertainty gained with a larger sample of SNe~Ia.  Thus, our
conclusions with only 16 SNe~Ia are already limited by systematic
uncertainties, {\it not} by statistical uncertainties.
  
   Here we explore possible systematic effects that might invalidate
our results. Of those that can be quantified at the present time, none
appears to reconcile the data with $\Omega_\Lambda = 0$.  Further
details can be found in Schmidt \etal ~(1998) and especially Riess 
\etal ~(1998b).

\subsubsection{Evolution}

   The local sample of SNe~Ia displays a weak correlation between
light-curve shape (or luminosity) and host galaxy type, in the sense
that the most luminous SNe~Ia with the broadest light curves only
occur in late-type galaxies.  Both early-type and late-type galaxies
provide hosts for dimmer SNe Ia with narrower light curves (Hamuy
\etal ~1996a).  The mean luminosity difference for SNe~Ia in late-type
and early-type galaxies is $\sim 0.3$ mag.  In addition, the SN~Ia
rate per unit luminosity is almost twice as high in late-type galaxies
as in early-type galaxies at the present epoch (Cappellaro \etal
~1997).  These results may indicate an evolution of SNe~Ia with
progenitor age. Possibly relevant physical parameters are the mass,
metallicity, and C/O ratio of the progenitor (H\"{o}flich \etal ~1998).

   We expect that the relation between light-curve shape and luminosity
that applies to the range of stellar populations and progenitor ages
encountered in the late-type and early-type hosts in our nearby sample
should also be applicable to the range we encounter in our distant
sample.  In fact, the range of age for SN Ia progenitors in the nearby
sample is likely to be {\it larger} than the change in mean progenitor
age over the 4--6 Gyr lookback time to the high-$z$ sample.  Thus, to
first order at least, our local sample should correct our distances
for progenitor or age effects.

   We can place empirical constraints on the effect that a change in the
progenitor age would have on our SN~Ia distances by comparing
subsamples of low-redshift SNe~Ia believed to arise from old and young
progenitors.  In the nearby sample, the mean difference between the
distances for the early-type (8 SNe~Ia) and late-type hosts (19
SNe~Ia), at a given redshift, is 0.04 $\pm$ 0.07 mag from the MLCS
method.  This difference is consistent with zero.  Even if the SN~Ia
progenitors evolved from one population at low redshift to the other
at high redshift, we still would not explain the surplus in mean
distance of 0.25 mag over the $\Omega_\Lambda=0$ prediction.  

  Moreover, it is reassuring that initial comparisons of high-redshift
SN~Ia spectra appear remarkably similar to those observed at low
redshift.  For example, the spectral characteristics of SN 1998ai ($z
= 0.49$) appear to be essentially indistinguishable from those of
low-redshift SNe~Ia (Riess \etal ~1998b).

   We expect that our local calibration will work well at eliminating
any pernicious drift in the supernova distances between the local and
distant samples. However, we need to be vigilant for changes in the
properties of SNe~Ia at significant lookback times.  Our distance
measurements could be especially sensitive to changes in the colors of
SNe~Ia for a given light-curve shape.

\subsubsection{Extinction}
 
   Our SN~Ia distances have the important advantage of including
corrections for interstellar extinction occurring in the host galaxy
and the Milky Way. Extinction corrections based on the relation
between SN~Ia colors and luminosity improve distance precision for a
sample of nearby SNe~Ia that includes objects with substantial
extinction (Riess \etal ~1996a); the scatter in the
Hubble diagram is much reduced.  Moreover, the consistency of the
measured Hubble flow from SNe~Ia with late-type and early-type hosts
(see above) shows that the extinction corrections applied to dusty
SNe~Ia at low redshift do not alter the expansion rate from its value
measured from SNe~Ia in low dust environments. 

   In practice, our high-redshift SNe~Ia appear to suffer negligible
extinction; their $B-V$ colors at maximum brightness are normal,
suggesting little color excess due to reddening. Riess \etal
~(1996b) found indications that the Galactic ratios between
selective absorption and color excess are similar for host galaxies in
the nearby ($z \leq 0.1$) Hubble flow.  Yet, what if these ratios
changed with lookback time?  Could an evolution in dust grain size
descending from ancestral interstellar ``pebbles'' at higher redshifts
cause us to underestimate the extinction?  Large dust grains would not
imprint the reddening signature of typical interstellar extinction
upon which our corrections rely.  

   However, viewing our SNe through such grey interstellar grains
would also induce a {\it dispersion} in the derived distances. Using
the results of Hatano \etal ~(1998), Riess \etal ~(1998b)
estimate that the expected dispersion would be 0.40 mag if the mean
grey extinction were 0.25 mag (the value required to explain the
measured MLCS distances without a cosmological constant).  This is
significantly larger than the 0.21 mag dispersion observed in the
high-redshift MLCS distances.  Furthermore, most of the observed
scatter is already consistent with the estimated {\it statistical}
errors, leaving little to be caused by grey extinction.  Nevertheless,
if we assumed that {\it all} of the observed scatter were due to grey
extinction, the mean shift in the SN~Ia distances would only be 0.05
mag.  With the observations presented here, we cannot rule out this
modest amount of grey interstellar extinction.

  Grey {\it intergalactic} extinction could dim the SNe without either
telltale reddening or dispersion, if all lines of sight to a given
redshift had a similar column density of absorbing material.  The
component of the intergalactic medium with such uniform coverage
corresponds to the gas clouds producing Lyman-$\alpha$ forest
absorption at low redshifts.  These clouds have individual H~I column
densities less than about $10^{15} \, {\rm cm^{-2}}$ (Bahcall \etal
~1996).  However, they display low metallicities, typically less than
10\% of solar. Grey extinction would require larger dust grains which
would need a larger mass in heavy elements than typical interstellar
grain size distributions to achieve a given extinction.  Furthermore,
these clouds reside in hard radiation environments hostile to the
survival of dust grains.  Finally, the existence of grey intergalactic
extinction would only augment the already surprising excess of
galaxies in high-redshift galaxy surveys (Huang \etal ~1997).

  We conclude that grey extinction does not seem to provide an
observationally or physically plausible explanation for the observed
faintness of high-redshift SNe~Ia.

\subsubsection{Selection Bias}

    Sample selection has the potential to distort the comparison of
nearby and distant SNe. Most of our nearby ($z < 0.1$) sample
of SNe~Ia was gathered from the Cal\'an/Tololo survey (Hamuy \etal
~1993), which employed the blinking of photographic plates obtained at
different epochs with Schmidt telescopes, and from less well-defined
searches (Riess \etal ~1998c).  Our distant ($z \gtrsim 0.16$) sample was
obtained by subtracting digital CCD images at different epochs with
the same instrument setup.

   Although selection effects could alter the ratio of intrinsically
dim to bright SNe~Ia in the nearby and distant samples, our use of the
light-curve shape to determine the supernova's luminosity should
correct most of this selection bias on our distance estimates.
Nevertheless, the final dispersion is nonzero, and to investigate its
consequences we used a Monte Carlo simulation; details are given by
Riess \etal ~(1998b).  The results are extremely encouraging, with
recovered values of $\Omega_M$ or $\Omega_\Lambda$ exceeding the
simulated values by only 0.02--0.03 for these two parameters
considered separately.  There are two reasons we find such a small
selection bias in the recovered cosmological parameters.  First, the
small dispersion of our distance indicator ($\sigma \approx 0.15$ mag
after light-curve shape correction) results in only a modest selection
bias.  Second, both nearby and distant samples include an excess of
brighter than average SNe, so the {\it difference} in their individual
selection biases remains small.

   Future work on quantifying the selection criteria of the nearby and
distant samples is needed.  Although the above simulation and others
bode well for using SNe~Ia to measure cosmological parameters, we must
continue to be wary of subtle effects that might bias the comparison
of SNe~Ia near and far.

\subsubsection{Effect of a Local Void}

   Zehavi \etal ~(1998) find that the SNe~Ia out to 7000 km s$^{-1}$
may (2--3$\sigma$ confidence level) exhibit an expansion rate which is
6\% greater than that measured for the more distant objects; see
Figure 1.  The implication is that the volume out to this distance is
underdense relative to the global mean density.

  In principle, a local void would increase the expansion rate measured
for our low-redshift sample relative to the true, global expansion
rate.  Mistaking this inflated rate for the global value would give
the false impression of an increase in the low-redshift expansion rate
relative to the high-redshift expansion rate.  This outcome could be
incorrectly attributed to the influence of a positive cosmological
constant. In practice, only a small fraction of our nearby sample is
within this local void, reducing its effect on the determination of
the low-redshift expansion rate.

   As a test of the effect of a local void on our constraints for the
cosmological parameters, we reanalyzed the data discarding the seven
SNe~Ia within 7000 km s$^{-1}$ ($d = 108$ Mpc for $H_0 = 65$ km
s$^{-1}$ Mpc$^{-1}$).  The result was a reduction in the confidence
that $\Omega_\Lambda > 0$ from 99.7\% (3.0$\sigma$) to 98.3\%
(2.4$\sigma$) for the MLCS method.

\subsubsection{Weak Gravitational Lensing}

  The magnification and demagnification of light by large-scale
structure can alter the observed magnitudes of high-redshift
SNe (Kantowski \etal ~1995).  The effect of weak
gravitational lensing on our analysis has been quantified by
Wambsganss \etal ~(1997) and summarized by Schmidt \etal ~(1998).  
SN~Ia light will, on average, be demagnified by $0.5$\% at $z=0.5$ and
$1$\% at $z=1$ in a Universe with a non-negligible cosmological
constant.  Although the sign of the effect is the same as the
influence of a cosmological constant, the size of the effect is
negligible.

   Holz \& Wald (1998) have calculated the weak lensing effects on
supernova light from ordinary matter which is not smoothly distributed
in galaxies but rather clumped into stars (i.e., dark matter contained
in massive compact halo objects).  With this scenario, microlensing
becomes a more important effect, further decreasing the observed
supernova luminosities at $z=0.5$ by 0.02 mag for $\Omega_M$=0.2 (Holz
1998).  Even if most ordinary matter were contained in compact
objects, this effect would not be large enough to reconcile the SNe Ia
distances with the influence of ordinary matter alone.

\subsubsection{Sample Contamination}

   Riess \etal ~(1998b) consider in detail the possibility of
sample contamination by SNe that are not SNe~Ia. Of the 16 objects,
12 are clearly SNe~Ia and 2 others are almost certainly SNe~Ia
(though the region near Si~II $\lambda$6150 was poorly observed
in the latter two). One object (SN 1997ck at $z = 0.97$) does not
have a good spectrum, and another (SN 1996E) might be a SN~Ic.
A reanalysis with only the 14 most probable SNe~Ia does not
significantly alter our conclusions regarding a positive
cosmological constant. However, without SN 1997ck we cannot
obtain independent values for $\Omega_M$ and $\Omega_\Lambda$. 

   Riess \etal ~(1998b) and Schmidt \etal ~(1998) discuss several
other uncertainties (e.g., differences between fitting techniques;
$K$-corrections) and suggest that they, too, are not serious in our
study.
 
\section{Conclusions}

   When used with care, SNe~Ia are excellent cosmological probes; the
current precision of individual distance measurements is roughly
5--10\%, but a number of subtleties must be taken into account to
obtain reliable results. SNe~Ia at $z \lesssim 0.1$ have been used to
demonstrate that (a) the Hubble flow is definitely linear, (b) the
value of the Hubble constant is $65 \pm 2$ (statistical) $\pm 5$
(systematic) km s$^{-1}$ Mpc$^{-1}$, (c) the bulk motion of the Local
Group is consistent with that derived from COBE measurements, and (d)
the dust in other galaxies is similar to Galactic dust.

   More recently, we have used a sample of 16 high-redshift SNe~Ia ($0.16
\leq z \leq 0.97$) to deduce the following.

(1) The luminosity distances exceed the prediction of a low
mass-density ($\Omega_M$ $\approx 0.2$) Universe by about 0.25 mag.  A
cosmological explanation is provided by a positive cosmological
constant at the 99.7\% (3.0$\sigma$) confidence level, with the prior
belief that $\Omega_M \geq 0$. We also find that the expansion of the
Universe is currently accelerating ($q_0 \leq 0$, where $q_0 \equiv
\Omega_M/2 - \Omega_\Lambda$), and that the Universe will expand
forever.

(2) The dynamical age of the Universe is 14.2 $\pm 1.7$ Gyr, including
systematic uncertainties in the Cepheid distance scale used for the
host galaxies of three nearby SNe~Ia.

(3) These conclusions do not depend on inclusion of SN 1997ck ($z =
0.97$; uncertain classification and extinction), nor on which of two
light-curve fitting methods is used to determine the SN~Ia distances.

(4) The systematic uncertainties presented by grey extinction, sample
selection bias, evolution, a local void, weak gravitational lensing,
and sample contamination currently do not provide a convincing
substitute for a positive cosmological constant.  Further studies
of these and other possible systematic uncertainties are needed to
increase the confidence in our results.

   We emphasize that the most recent results of the SCP are consistent
with those of our HZT. This is reassuring --- but other, completely
independent methods are certainly needed to verify these conclusions. 
The upcoming space CMBR experiments hold the most promise
in this regard. Although new questions will undoubtedly arise (e.g.,
What is the physical source of the cosmological constant, if nonzero?
Are evolving cosmic scalar fields a better alternative?), we speculate
that this may be the beginning of the ``end game" in the quest for the
values of $\Omega_M$ and $\Omega_\Lambda$.

\bigskip

  We thank all of our collaborators on the HZT for their contributions
to this work. Our supernova research at U.C. Berkeley is supported by
the Miller Institute for Basic Research in Science, by NSF grant
AST-9417213, and by grant GO-7505 from the Space Telescope Science
Institute, which is operated by the Association of Universities for
Research in Astronomy, Inc., under NASA contract NAS5-26555. A.V.F. is
grateful to the Committee on Research at U.C. Berkeley for travel
funds to attend this meeting.

\medskip

\centerline {FIGURE CAPTIONS}

\medskip
 
% {\bf Figure 1:} Light curves of high-redshift SNe~Ia.  $B$ (filled
% symbols) and $V$ (open symbols) photometry in the rest frame of 10
% well-observed SNe~Ia is shown with $B$ increased by 1 mag for ease
% of view.  The lines are the empirical MLCS model fits to the
% data.  Supernova age is shown relative to $B$ maximum.

% Riess et al. Figure 4

{\bf Figure 1:} The upper panel shows the Hubble diagram for the
low-redshift and high-redshift SN~Ia samples with distances measured
from the MLCS method. Overplotted are three world models: ``low'' and
``high'' $\Omega_M$ with $\Omega_\Lambda=0$, and the best fit for a
flat universe ($\Omega_M=0.24$, $\Omega_\Lambda=0.76$).  The bottom
panel shows the difference between data and models with
$\Omega_M=0.20$, $\Omega_\Lambda=0$.  The open symbol is SN 1997ck
($z=0.97$) which lacks spectroscopic classification and an extinction
measurement.  The average difference between the data and the
$\Omega_M=0.20$, $\Omega_\Lambda=0$ prediction is 0.25 mag.

\medskip

% Riess et al. Figure 6

{\bf Figure 2:} Joint confidence intervals for
($\Omega_M$,$\Omega_\Lambda$) from SNe~Ia.  The solid contours are
results from the MLCS method applied to well-observed SN~Ia light
curves, together with the snapshot method (Riess \etal 1998a) applied
to incomplete SN~Ia light curves.  The dotted contours are for the
same objects excluding SN 1997ck ($z=0.97$).  Regions representing
specific cosmological scenarios are illustrated.  

% {\bf Figure 4:} Spectral comparison (in f$_{\lambda}$) of SN 1998ai
% ($z=0.49$) with low-redshift ($z < 0.1$) SNe~Ia at a similar age
% ($\sim 5$ days before maximum brightness).  Within the narrow range of
% SN~Ia spectral features, SN 1998ai is indistinguishable from the
% low-redshift SNe~Ia.  The spectra of the low-redshift SNe~Ia were
% resampled and convolved with Gaussian noise to match the quality of
% the spectrum of SN 1998ai.
 
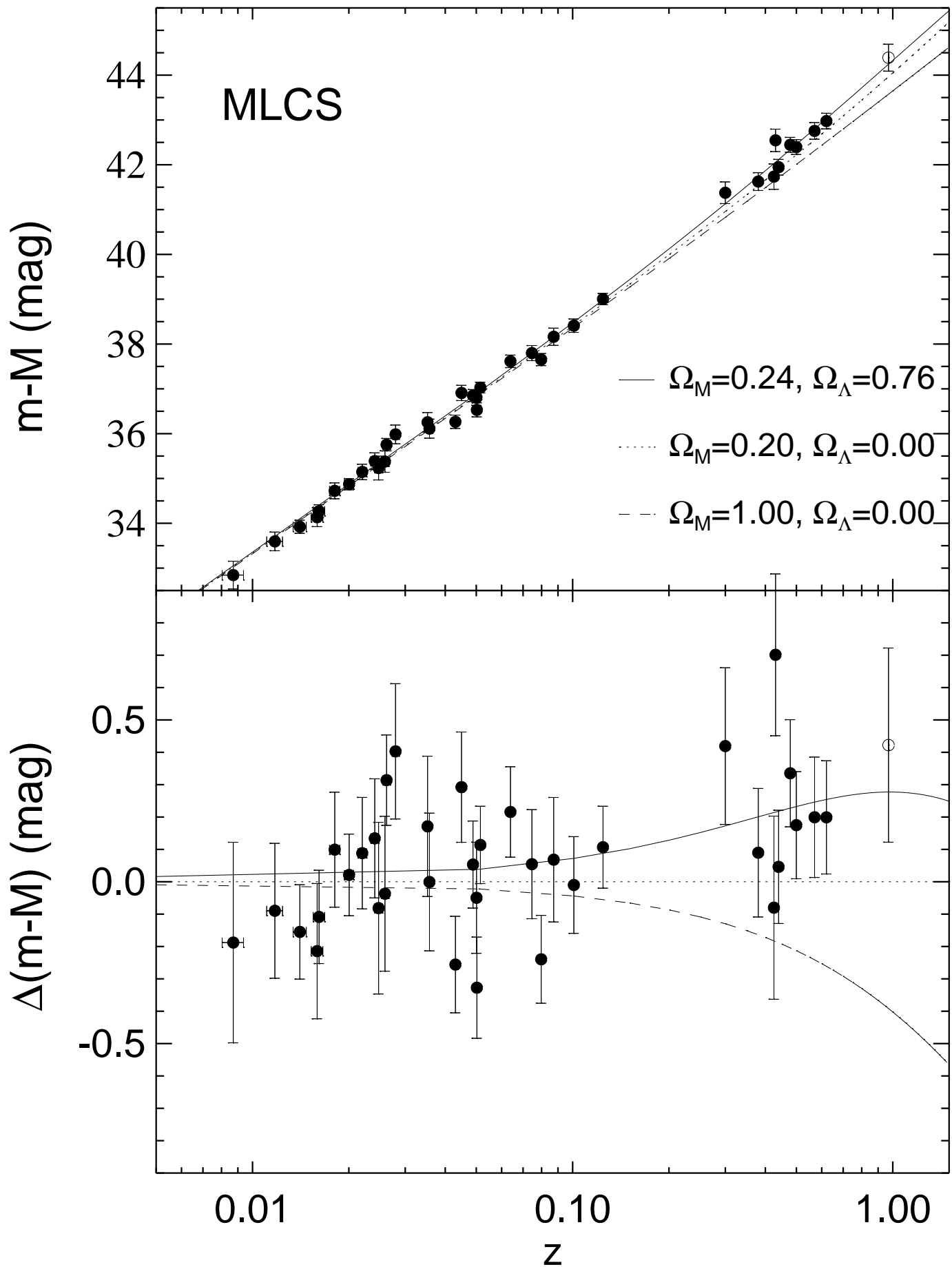
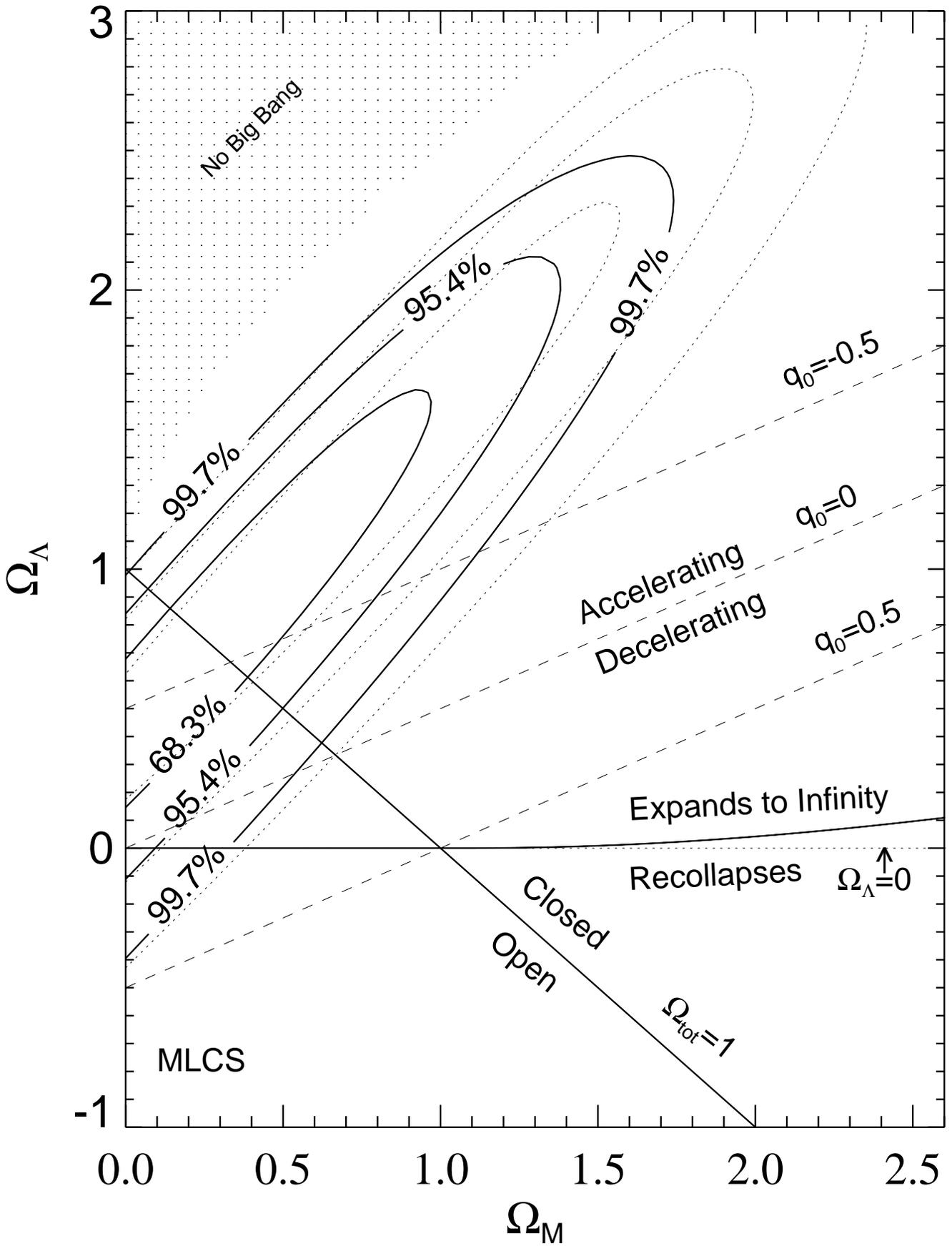
\end{document}